\documentstyle[manuscript,aps]{revtex}
\tighten
\begin{document}
\draft
\tolerance = 10000
\hfuzz=5pt
\preprint{\vbox{To appear in Physical Review {\bf D}\hfill
         USC(NT)-98-02}}
\tolerance = 10000
\hfuzz=5pt\title{Reply to Comment on
``Hara's theorem in the constituent quark model"}
\author{V.~Dmitra\v sinovi\' c}
\address{  
Department of Physics and Astronomy, \\
University of South Carolina,
Columbia, SC 29208, USA 
\\Email address:  dmitra@tor.physics.sc.edu}
\maketitle
\begin{abstract}
In the preceding Comment it is alleged that a 
``hidden loophole'' in the proof of Hara's theorem has been found, which 
purportedly invalidates the conclusions of the paper commented upon.
I show that there is no such loophole in the constituent quark model, 
and that the ``counterexample'' presented in the Comment is not gauge invariant.
\end{abstract}
\pacs{PACS numbers: 11.30.Hv, 13.30.-a, 14.20.Jn}
\widetext

In the Comment \cite{z98} on my paper ``Hara's theorem in the
constituent quark model" \cite{vd96},
it is pointed out that the ``hidden assumption'' of sufficiently
localized current is tacitly made in my argument and then claimed that 
this assumption is not valid in the constituent quark model. 

In the following I show that: (a) this assumption is well known, 
(b) the mathematical formulation of this assumption as used in the Comment 
\cite{z98} is incorrect,
(c) the correct form of this assumption is satisfied by the constituent quark 
model, and  (d) the electromagnetic (EM) current 
conservation is violated in the constituent quark model calculation
of Kamal and Riazuddin (KR) \cite{kr83} which current is used as 
a ``counterexample'' in Ref. \cite{z98}. 

\paragraph*{(a)}
That ``a sufficiently localized current" is a well known 
condition for the existence of the multipole expansion,
can be seen in textbooks, see for example p. 54 of Ref.
\cite{wal95}, [between Eqs. (7.23) and (7.24)], 
where it emerges from the demand that all surface terms vanish in integrations
by parts leading to the multipole expansion of the EM Hamiltonian
matrix element (ME)
\begin{eqnarray}
H_{fi} &=& H_{fi}({\bf q}) = 
- \int d{\bf R}~\hat{\bbox{\varepsilon}}_{M} \cdot 
{\bf J}_{fi}({\bf R}) ~\exp(i{\bf q} \cdot {\bf R}) 
\nonumber \\
&=&  
\sum_{J = 1}^{\infty} \sqrt{2 \pi (2J + 1)} i^{J} 
\left[ \lambda
\langle \Psi_{f}|\hat{T}_{JM}^{\rm mag}| \Psi_{i} \rangle 
+ 
\langle \Psi_{f}|\hat{T}_{JM}^{\rm el}| \Psi_{i} \rangle
\right],
\label{e:multi} \
\end{eqnarray}
where $M = \pm$ and $q = |{\bf q}|$. Equivalently, the
assumption is that the EM Hamiltonian ME $H_{fi}({\bf q})$ 
itself be a well-defined (proper) integral. 

\paragraph*{(b)}
The mathematical formulation of the ``implicit assumption" that
was offered in Ref. \cite{z98} reads   
\begin{equation}
{\bf J}({\bf R}) \cdot \hat{\bbox{\varepsilon}}_{+}  < R^{-3},
~~~{\rm as}~~~ R = |{\bf R}| \rightarrow \infty. 
\label{ineq}
\end{equation}
Strictly speaking this inequality is meaningless, since it compares an
operator valued (q-number) left-hand side with an ordinary (c-number)
function right-hand side.
If we accept this inequality as a statement about matrix elements then we 
can show that the conclusions drawn in Ref. \cite{z98} do {\it not} follow 
from it. Specifically, this inequality, although violated by the
``counterexample" transverse current in Eq. (7) of 
Ref. \cite{z98}, is actually {\it not} the 
source of the unusual threshold behavior in 
$\hat{T}_{1M}^{\rm el}$, Eqs. (8), (12) in Ref. \cite{z98}. 
Rather, the real ``culprit" is the singular 
behavior of ${\bf J}(R) \cdot \hat{\bbox{\varepsilon}}_{+}$ as 
$R \to 0$.

To prove this assertion note that the EM Hamiltonian $H_{fi}({\bf q})$, 
Eq. (\ref{e:multi}),
with the ``counterexample'' EM current defined by Eq. (4) in Ref. \cite{z98}, 
is infinite as its stands.
That fact alone should have been enough to suggest that the subsequent 
conclusions would be questionable. To better define this integral
a ``regularization" procedure was introduced into Eq. (4) of Ref. \cite{z98} 
in the form of a Gaussian [as a function of the new parameter $\varepsilon$] 
multiplying the integrand. 
But, rather than keeping the Gaussian regularization until the 
end of the calculation, it was removed too soon.
This procedure led to the erroneous conclusions drawn in Ref. \cite{z98}. 
Specifically, the $\varepsilon \to 0$ limit is taken under the 
integral sign in Eq. (11) in Ref. \cite{z98}, the relevant part being
\begin{eqnarray}
\alpha &=&
\lim_{\varepsilon \to 0} 
q \int_{0}^{\infty} dr j_{1}(qr) {\rm erf}\left({r \over{2
\sqrt{\varepsilon}}}\right) =
\int_{0}^{\infty} dz j_{1}(z) \ . 
\label{int1}
\end{eqnarray}
A more careful calculation of the integral (\ref{int1}) leads to
\begin{eqnarray}
\alpha &=&
\left({2 \over{q}}\right)
\int_{0}^{\infty} dz j_{0}(z) 
\delta \left({z \over{q}}\right)  \ . 
\label{int3}
\end{eqnarray}
Note that the integral (\ref{int3}) receives its 
whole value from only one point - the lower integration bound $r=0$ - 
and not from $r \to \infty$, as implied by inequality (\ref{ineq})
and claimed in Ref. \cite{z98}. 

Manifestly an object with a Dirac delta function singularity is localized. 
[This is not a proof that the physical EM current hyperon matrix element (ME) 
is localized--that will be checked in the next section]. 
So, currents of the type of Eq. (4) in Ref. \cite{z98}, 
if they exist, are a new and (very) short-distance 
phenomenon. They must come from some high-energy extension of the 
Standard Model, since they do not exist in the Salam-Weinberg (SW) model 
\footnote{The SW model determines the form of the electroweak currents in the 
constituent quark model.}. 
For single-quark current operators this can be seen from the relevant SW
Feynman rules. It is alleged in Ref. \cite{z98} that the effective 
two-quark EM current operators [induced 
by the $W^{\pm}, Z^0$ exchange graphs] form such an ``abnormal'' current.
We shall show in point $(d)$ below that this claim is incorrect because the said
current is {\it not} conserved. But, first we shall show that the EM current 
hyperon matrix element in the constituent quark model is sufficiently localized to 
have a normal threshold behavior.  

\paragraph*{(c)}
As already stated in point $(a)$ above, the sufficient condition for 
localizability is the existence of the Fourier transform (FT) in Eq. 
(\ref{e:multi}), i.e., at least absolute-integrability ($\alpha = 1$) 
\footnote{Absolute integrability ensures the existence of the (direct) Fourier
transform, but not that of the inverse one. To ensure the
existence of both, one needs square-integrability ($\alpha = 2$).} 
\cite{mess59} of ${\bf J}_{fi}({\bf R}) \cdot \hat{\bbox{\varepsilon}}_{+}$:
\begin{equation}
\int d{\bf R} |{\bf J}_{fi}({\bf R}) \cdot 
\hat{\bbox{\varepsilon}}_{+}|^{\alpha} = 
\int d{\bf R} |\langle \Psi_{f}| 
{\bf J}({\bf R}) \cdot \hat{\bbox{\varepsilon}}_{+}|\Psi_{i} 
\rangle |^{\alpha} < \infty ~; ~~\alpha = 1, 2. 
\label{abs}
\end{equation}
Hence we see that the hyperon wave functions $|\Psi_{i, f} \rangle$ play 
an important role in deciding localizability. 
The complete hyperon wave function $|\Psi \rangle$ factors into the 
center-of-mass (CM) 
plane wave and the internal (quark) wave function (w.f.) $|\Phi \rangle$
\begin{equation}
| \Psi_{i}({\bf P}_{i}) \rangle \simeq 
\exp \left(i ({\bf P}_{i} \cdot {\bf X} - E_{i} t)\right)
|\Phi (\bbox{\rho},\bbox{\lambda}) \rangle~~. 
\label{cmwf}
\end{equation}
where $(E_{i}, {\bf P}_{i})$ is the initial-state hyperon four-momentum and 
$\bbox{\rho}, \bbox{\lambda}$ are
the three-body Jacobi coordinates describing the motion of the three
constituent quarks in the hyperon relative to its CM co-ordinate $\bf X$. 
The internal wave functions $|\Phi_{i,f} \rangle$ 
are bound-state ones and therefore normalizable, whereas
the CM plane waves produce a (non-square-integrable) 
momentum-conserving Dirac delta function in all momentum space matrix elements, 
\begin{eqnarray}
- H_{fi}({\bf q}) 
&=&  
\int d{\bf R}~\hat{\bbox{\varepsilon}}_{+} \cdot 
{\bf J}_{fi}(R) ~\exp(i{\bf q} \cdot {\bf R}) 
\nonumber \\
&=& 
(2 \pi)^{3} \delta ({\bf P}^{'} - {\bf q} - {\bf P})~
\int d{\bf R}~\langle \Phi_{f}| 
{\bf J}({\bf R}) \cdot \hat{\bbox{\varepsilon}}_{+}|\Phi_{i} 
\rangle ~\exp(i{\bf q} \cdot {\bf R})~, 
\label{e:me2} \
\end{eqnarray}
which we systematically dropped from the displayed equations in our 
previous publications.
Thus we see that the original integral in the criterion Eq. (\ref{abs}) 
is singular due to this trivial CM motion. It is the remaining (form) factor
$\int d{\bf R}~\langle \Phi_{f}| 
{\bf J}({\bf R}) \cdot \hat{\bbox{\varepsilon}}_{+}|\Phi_{i} 
\rangle ~\exp(i{\bf q} \cdot {\bf R})$, with ``the CM motion taken out'' 
that must be a well-defined Fourier transform: 
\begin{equation}
\int d{\bf R} |\langle \Phi_{f}| 
{\bf J}({\bf R}) \cdot \hat{\bbox{\varepsilon}}_{+}|\Phi_{i} 
\rangle | < \infty , 
\label{abs1}
\end{equation}
which is the final form of the localizablity criterion.
The physical meaning of this requirement is clear: 
the hyperon probability distribution weighted by the EM current must 
be sufficiently close to its CM to yield a finite expectation value integral.
This condition is satisfied by the hyperon in the constituent quark model
because of the normalizability of its internal (quark) wave functions 
$|\Phi_{i,f} \rangle$;
confinement makes the quark probability density only more localized
and the integral faster converging. The EM current
operator is at most a polynomial in the momenta (gradients) and spin operators
which cannot overcome the exponential decay of the internal wave functions. 
Consequently the hyperon EM current ME is localized in the constituent quark 
model, as advertised. 

\paragraph*{(d)}
Finally we turn to the question of EM current conservation in the 
constituent quark model calculation of Kamal and Riazuddin \cite{kr83},
which is used as a ``counterexample'' 
in Ref. \cite{z98}.
First note that there is full agreement between
the KR paper \cite{kr83} and 
Refs. \cite{vd96,vd92} on the question
of gauge invariance of the {\it covariant} amplitude
described by the Feynman diagrams in Figs. 1, 2 of 
Ref. \cite{kr83} (c.f. figures 1, 2  in Ref. \cite{vd92}) . 
However, {\it this does not mean that the result of its 
nonrelativistic reduction is also gauge invariant}. 
The nonrelativistic reduction of the parity violating part of
the Feynman amplitude leads KR to the current
\begin{equation}
{\bf J} \simeq \left(\bbox{\sigma}_{1} \times \bbox{\sigma}_{2}\right), 
\label{2-b}
\end{equation}
shown in Eq. (13) of Ref. \cite{kr83}. 
The EM current (\ref{2-b}) is only the two-body part of the complete EM current. 
By itself it is {\it not} conserved, see Eq. (15) in Ref. \cite{vd96}, or 
Eqs. (3.6-9) in Ref. \cite{vd92}. 
To convince oneself explicitly of this fact, compare the EM 
current Eq. (\ref{2-b})
with the corresponding manifestly conserved current, 
\begin{equation}
{\bf J} \simeq \hat{\bf q} \times 
\left(\left(\bbox{\sigma}_{1} 
\times \bbox{\sigma}_{2}\right) \times \hat{\bf q}\right)
= 
\left(\bbox{\sigma}_{1} \times 
\bbox{\sigma}_{2}\right)
- \hat{\bf q} \left(\bbox{\sigma}_{1} \times 
\bbox{\sigma}_{2}\right) \cdot \hat{\bf q}. 
\label{2-b1}
\end{equation}
One can see that the current (\ref{2-b}) is just the first term
on the right-hand side of Eq. (\ref{2-b1}) - there is no term 
proportional to the 
three-momentum transfer $\hat{\bf q}$. Consequently the  
current (\ref{2-b}) is {\it not} transverse to 
$\bf q$ in momentum space, i.e., {\it it is not conserved}. 
Since the two-body term constitutes the whole EM current in 
the KR calculation, we have proven our last contention: that the 
calculation in Ref. \cite{kr83} is {\it not} gauge invariant. 

Thus we have shown that the objections raised in Ref. \cite{z98}
are invalid. 

\paragraph*{Acknowledgments}
The author would like to thank F. Myhrer and K. Kubodera for
reading the manuscript and for valuable comments.

\end{document}